# Security proof of the two-way quantum secure direct communication with channel loss and noise

Jianyong Hu, Mingyong Jing, Peng Zhang, Qiangqiang Zhang, Huifang Hou, Liantuan Xiao, and Suotang Jia

*State Key Laboratory of Quantum Optics and Quantum Optics Devices, Institute of Laser Spectroscopy, Shanxi University, Taiyuan 030006, China*

*Collaborative Innovation Center of Extreme Optics, Shanxi University, Taiyuan 030006, China;*

*Corresponding author E-mail address: xlt@sxu.edu.cn

Quantum secure direct communication is one of the important mode of quantum communication, which sends secret information through a quantum channel directly without setting up a prior key. Over the past decade, numerous protocols have been proposed, and part of them have been experimentally demonstrated. However, the strict security proof is still in suspense. Two-way protocol is seen as one of the most practical protocol, in this paper, the security of the two-way quantum secure direct communication protocol will be proved against general attacks when the noisy and lossy channel is taken into account. There are two steps included in the proof, firstly, we need to prove that the mutual information between Alice and Bob is larger than the mutual information shared between Alice and the eavesdropper when the error rate in control mode is less than the predetermined threshold. Secondly, once the error rate is lower than the threshold, one can always find out a forward error correction code to guarantee the security of information transmission.

## I. INTRODUCTION

Quantum communication enables two remote parties to share secret information securely over a long distance [1]. Since the pioneering protocol was presented by Bennett and Brassard [2], different modes of quantum communication have been developed, such as quantum key distribution (QKD), quantum secret sharing, quantum secure direct communication (QSDC), quantum teleportation, quantum dense coding, and so on [2-6].

QSDC is one of the important mode of the quantum communication, in contrast to QKD, QSDC sends secret information directly through a quantum channel with unconditional security without setting up a prior key [7]. Since the first QSDC protocol was proposed [4], it has become one of the hot research topics in quantum communication over the past decade [8, 9]. To the entanglement carriers, in 2003, Deng, Long and Liu proposed the two-step QSDC protocol where the criteria for QSDC were explicitly stated [10]. QSDC protocols based on high-dimensional entanglement, multipartite entanglement, and hyperentanglement were developed [11-14]. To the single photons carriers, the first QSDC protocol was proposed in Ref. [15], the so-called DL04 protocol, its feasibility have been demonstrated [16-18]. Wei Zhang .et al carried out QSDC experiment with quantum memory [19]. In addition, protocols of quantum signature, quantum dialogues, and quantum direct secret sharing have been constructed based on QSDC [20-22].

In the practice, the channel loss and noise may lead to the error of the information bit when it is encoded on the individual photon [23]. Therefore, QSDC protocol uses a block transmission technique was proposed by Long and Liu, in which the quantum information carriers such as single photons or Einstein-Podolsky-Rosen entanglement pairs are transmitted in a block of *N* particles [4]. However, when there is noise in the quantum channel, an adversary Eve can gain a certain amount of qubits by hiding her presence in the channel noise. The information leakage may be eliminated by using quantum privacy amplification [24]. Unfortunately, quantum privacy amplification ruins the direct communication picture as it involves the merger and order reshuffling of qubits.

An efficient way to implement QSDC in noisy and lossy channel is to use classical forward error correction (FEC) code. Two-way protocol is seen as one of the most practical protocol [25-30]. In the previous work [16,31], a FEC code, named frequency coding scheme, was used in the two-way QSDC to overcome the channel loss and noise, however, the strict security proof of the protocol is still an outstanding question. In this work, we present a security proof of the two-way QSDC protocol against general attacks. There are two steps in the proof, we first prove that secure qubits is not zero when the error rate in control mode is lower than the predetermined threshold. Secondly, once the first condition is satisfied, one can always find out a FEC coding scheme which can transmit information securely.

## II. SECURITY OF TWO-WAY QSDC PROTOCOL

Suppose that Alice wants to send secret information to Bob. The two-way QSDC protocol works as follows:

(1) Bob prepares a block of $N_e$ single photons. Each photon in the block is randomly in one of the four states $|0\rangle$, $|1\rangle$, $|+\rangle$, and $|-\rangle$, where $|0\rangle$ and $|1\rangle$ are the eigenstates of the Pauli **Z** operator, and $|\pm\rangle=(|0\rangle\pm|1\rangle)/2$ are the eigenstates of the Pauli **X** operator. Then Bob sends the single-photon block to Alice, Alice acknowledges this fact.

(2) Because of channel noise and loss, Alice receives only $N_r$ single photons ($N_r=N_e\eta_f$), where $\eta_f$ is the transmission efficiency of Bob-Alice channel i.e. forward channel. He selects $CN_r$ number ($C$ is a positive number less or equal to 1/2) of photons randomly from the $N_r$ received photons for eavesdropping check by measuring them randomly in the $X$-basis or the $Z$-basis (Control mode). Then Alice tells Bob the positions, the measuring-basis and the measuring results of these measured photons. Bob compares his results with those of Alice and obtains an error rate $e$. If the error rate is higher than the threshold, they will abort the communication. If the error rate is less than the threshold, the Bob-Alice communication is considered as safe, and then continue to the next step.

(3) The remaining $(1-C)N_r$ received photons are used for encoding secret information (Encode mode). She also selects $C(1-C)N_1$ single photons from the remaining photons randomly as check bits for the Alice-Bob transmission, and applies randomly one of the two operations, $U=i\sigma_y=|0\rangle\langle 1|-|1\rangle\langle 0|$, and $I=|0\rangle\langle 0|+|1\rangle\langle 1|$, which flip or not flip the state of the photon. The rest of the single photons will be encoded for information transmission, the coding scheme is pre-negotiated and proper designed according to the error rate measured in the control mode.

(4) Alice sends the encoded photon block back to Bob who can deterministically decode the qubits by using the same measuring basis when he prepared the single photon. Bob gets the bit value of each single photon in the block and their arrival time. Because of channel loss, Bob receives only $N$ (here $N=(1-C)^2N_r\eta_b$, $\eta_b$ is the transmission efficiency of Alice-Bob channel i.e. backward channel) photons in each block after subtracting the check photons. Alice and Bob will also publicly compare the results of the checking bits to ensure if there exists eavesdropping in the Alice-Bob transmission. Next, Bob decodes the secret information from the block of qubits.

The security proof of the above QSDC protocol includes two steps, in the next section, we will show that if the error rate $e$ is less than the preset threshold $e_t$, secure qubit rate would bigger than zero, i.e. $I_{AB}>I_{AE}$, where $I_{AB}$ is the mutual information between Alice and Bob, $I_{AE}$ is the mutual information between Alice and eavesdropper (Eve). Then, we will prove once $I_{AB}>I_{AE}$, Alice can always find out a FEC code scheme to make the information transmission securely.

## A. Security against general attacks

The security analysis in this section draw on the work in Ref. [32-35]. At first, Bob prepares a block of qubits which randomly in the state $|0\rangle$, $|1\rangle$, $|+\rangle$, $|-\rangle$ with the same probability, i.e., the qubits are prepared in a mixed state, $\rho^B=(|0\rangle\langle 0|+|1\rangle\langle 1|+|+\rangle\langle +|+|-\rangle\langle -|)/4=(|0\rangle\langle 0|+|1\rangle\langle 1|)/2$. After Alice's encoding operation, to gain Alice's information bits, Eve has to distinguish $\rho_0^A=I\rho^B I$ from $\rho_1^A=U\rho^B U$ in the Alice-Bob channel. Since $\rho_0^A=\rho_1^A=(|0\rangle\langle 0|+|1\rangle\langle 1|)/2$, Eve cannot gain any information about Alice's information bits if she only attacks the qubits after Alice's encoding operation. Eve has to attack the qubits traveling on both the Bob-Alice channel and the Alice-Bob channel in order to gain Alice's information bits.

Eve's most general quantum operation can be described by a unitary operation together with an ancilla. Eve performs a collective attack, i.e., $U_{BE}$ are the same for all qubits.

$$U_{BE}|0\rangle_B|E\rangle = c_{00}|0\rangle_B|E_{00}\rangle + c_{01}|1\rangle_B|E_{01}\rangle$$
$$U_{BE}|1\rangle_B|E\rangle = c_{11}|1\rangle_B|E_{11}\rangle + c_{10}|0\rangle_B|E_{10}\rangle$$
$$U_{BE}|+\rangle_B|E\rangle = c_{++}|+\rangle_B|E_{++}\rangle + c_{+-}|-\rangle_B|E_{+-}\rangle$$
$$U_{BE}|-\rangle_B|E\rangle = c_{--}|-\rangle_B|E_{--}\rangle + c_{-+}|+\rangle_B|E_{-+}\rangle \quad (1)$$

Where $c_{ij}$, ($i,j=0, 1, +, -$) can be treated as nonnegative real number, and $|E_{ij}\rangle$, ($i,j=0, 1, +, -$) forms four pairs of normalized vectors. After Eve's attack in the Bob-Alice channel, the joint state of the forward qubit and Eve's ancilla becomes

$$\rho_{Bob-Alice}^{BE} = U_{BE}(\rho^B \otimes |E\rangle\langle E|)U_{BE}. \quad (2)$$

After receiving the forward qubits, in the control mode, Alice randomly select and measures the qubits with probability $C$ for attack-detection. By publishing the measurement results, the error rate $e$ of qubit could be obtained. In the encoding mode, Alice encode $k$ information bits onto the block of $N$ forward qubits. With probability $P_1$ and $P_0$, Alice encodes information bit 1 and 0 by the operation $I$ and $Y$, respectively. The state of the qubits and Eve's ancillas becomes

$$\rho^{ABE} = P_0|0\rangle\langle 0|^A \otimes \rho_0^{BE} + P_1|1\rangle\langle 1|^A \otimes \rho_1^{BE}, \quad (3)$$

where $\rho_0^{BE} = \rho_{Bob-Alice}^{BE}$, $\rho_1^{BE} = Y\rho_{Bob-Alice}^{BE}Y$. Then Alice sends the encoded qubits back to Bob. The quantum efficiency of the information transmission is defined as $r_s=\lim_{m\to\infty}k/n$, $n$ is the block size of encoded pulses, $k$ is the secure qubits of the block. In the asymptotic scenario, the secure qubit rate $r_s$ is bounded by the conditional entropy of information bits that Alice sends to Bob given the quantum information of Eve, $r_s=S(\rho^A|\rho^{BE})$. After Alice's encoding operation, Eve can gain some quantum information about Alice's qubit from the quantum state $\rho^{BE}=tr_A\rho^{ABE}$ that is a joint state of the backward qubit and her ancilla. Here, we assume the worst case that Eve uses the entire state $\rho^{BE}$ to gain information bit, even though she may have to send part of the state to Bob. $r_s=S(\rho^A|\rho^{BE})=S(\rho^{ABE})-S(\rho^{BE})$, here, $S(\rho^{ABE})=-\text{tr}\rho^{ABE}\log_2\rho^{ABE}$, $S(\rho^{BE})=-\text{tr}\rho^{BE}\log_2\rho^{BE}$. In the following, we should calculate the eigenvalues of $\rho^{ABE}$ and $\rho^{BE}$ to get $S(\rho^{ABE})$ and $S(\rho^{BE})$. According to Ref. [], the secure qubit rate

$$r_s = h(P_0) - h(\xi), \tag{4}$$

where $\xi = c_{++}^2 - c_1^2$, $P_0$ is the probability that Alice encode information bit 0 on the qubit, and $h(x)=-x\log_2 x-(1-x)\log_2(1-x)$ is the binary Shannon entropy.

For the practical quantum channels, channel loss and noise should be considered. Alice and Bob should estimate the error rate in the Bob-Alice channel, we assume that the error rate in the backward channel is the same as the error rate in the forward channel. In the asymptotic scenario, after verifying

$$e < 1/4, \tag{5}$$

Alice can sends the secret information against collective attacks with secure qubit rate

$$r_s = I_{AB} - I_{AE} = h(P_0) - h(\xi) - h(e) - \eta_b, \tag{6}$$

where $\eta_f$ is the channel loss in the backward channel. $I_{AB}$ is the mutual information between Alice and Bob, and $I_{AE}$ is the mutual information between Alice and Eve. As you can see, the final secure qubit rate relates not only to the error rate but also the coding scheme and channel loss.

### B. Secure information bit rate

In this part, we will prove that once the secret information transmission rate $r_s>0$, Alice can always find a coding scheme to transmit information securely. Assume there is a binary asymmetric channel, the channel loss is $\eta$, and error rate is $p_1$. When the photon is detected

$$\begin{aligned}p(0|0)&=1-p_1,\\p(1|0)&=p_1,\\p(0|1)&=p_1,\\p(1|1)&=1-p_1,\end{aligned} \tag{7}$$

when the photon is lost

$$p(0|0 \text{ or } 1) = p(1|0 \text{ or } 1) = 1/2. \tag{8}$$

Therefore, the binary entropy can be expressed as

$$\begin{aligned}H(p) &= (1-\eta_A)[-p_1\log p_1 - (1-p_1)\log(1-p_1)]\\&+\eta_A(-p_2\log p_2 - (1-p_2)\log(1-p_2))\\&=(1-\eta)[-p_1\log p_1 - (1-p_1)\log(1-p_1)] + \eta,\end{aligned} \tag{9}$$

where $p_2=1/2$ signify the loss of the qubit.

An $n$ bits sequence can be seen as a point in $n$-dimensional Hilbert space, at the receiving end, it change to a 'sphere' because of channel loss and noise. The number of point in the 'sphere' is

$$2^{nH(p)} = 2^{n(-p_1\log p_1 - (1-p_1)\log(1-p_1)+\eta)}. \tag{10}$$

In order to decode correctly, the Hamming distance of each code word should be bigger than $2^{nH(p)}$, in addition,

$$2^{nH(p)} \cdot 2^{nR} = 2^{n\{(1-\eta)[-p_1\log p_1-(1-p_1)\log(1-p_1)]+\eta\}} \cdot 2^{nR} \leq 2^n \tag{11}$$

That is

$$0 \leq R \leq C(p) \equiv 1 - H(p), \tag{12}$$

where $R$ is the transmission efficiency, when $R<C(p)$ one can always find a coding scheme to transmit information correctly according to Shannon's noisy-channel coding theorem.

Assume that $k$ ($k<n$) information bits are encoded on the $n$ qubits. Because of channel loss and noise, part of qubits can be leaked to Eve as we discussed above. Therefore, Eve can be seen as a receiver whose loss rate is $\eta_E$, which means Eve gains $n(1-\eta_E)$ qubits, here the loss rate is originated from the uncertainty principle and control mode. The loss rate of legitimate receiver Bob is $\eta_B$, which is caused by channel loss and imperfect detection efficiency.

The error code is caused by channel noise and imperfect encoding operation of Alice, to consider the worst case, Eve use a perfect channel without noise and loss. The error rate caused by the channel loss and Alice's encoding operation is $p_c$ and $p_A$ respectively.

The code word is expanded to a Hamming sphere because of the channel loss and noise. The number of typical code words of Eve is

$$2^{n\{(1-\eta_E)[-p_A\log p_A-(1-p_A)\log(1-p_A)]+\eta_E\}}. \tag{13}$$

The number of typical code words of Bob is

$$2^{n\{(1-\eta_A)[-p_1\log p_1-(1-p_1)\log(1-p_1)]+\eta_A\}}, \tag{14}$$

where $p_1=p_c+p_A$.

The Hamming distance of each typical code word that Alice used must bigger than the radius of the Hamming sphere. If the Hamming distance of the detected code word and the typical code word is bigger than the radius of the Hamming sphere, error would be happened. The following two condition must be satisfied

$$2^{n\{(1-\eta_B)[-p_1\log p_1-(1-p_1)\log(1-p_1)]+\eta_B\}} \cdot 2^{nR} \leq 2^n, \tag{15a}$$

$$\frac{2^{n\{(1-\eta_E)[-p_A\log p_A-(1-p_A)\log(1-p_A)]+\eta_E\}}}{2^{n\{(1-\eta_B)[-p_1\log p_1-(1-p_1)\log(1-p_1)]+\eta_B\}}} > 2^2. \tag{15b}$$

The condition (15a) keep the information transmission accurately, the condition (15b) assure the security of information transmission.

When the radius of Eve's Hamming sphere is three times of Bob's, Eve can't distinguish adjacent typical code words, here the adjacent typical code words means the typical code words which has closest Hamming distance. Although, the maximum transmission capacity is $nR$, to guarantee the security of information transmission, only one information bit could be encoded on the block, and the information bit should be encoded on the adjacent typical code words. With the increase of the radius of Eve's Hamming sphere, more information bits could be encoded on the block. The final information transmission efficiency is

$$\begin{aligned}r &= \log[\frac{2^{(1-\eta_E)[-p_A\log p_A-(1-p_A)\log(1-p_A)]+\eta_E}}{2^{(1-\eta_B)[-p_1\log p_1-(1-p_1)\log(1-p_1)]+\eta_B}}]\\&=\{(1-\eta_E)[-p_A\log p_A-(1-p_A)\log(1-p_A)]+\eta_E\}\\&\quad-\{(1-\eta_B)[-p_1\log p_1-(1-p_1)\log(1-p_1)]+\eta_B\}.\end{aligned} \tag{16}$$

Figure 1 shows the distribution of information transmission efficiency changes with $\eta_E$ and $\eta_B$. Only when $\eta_E>\eta_B$, the secure encoding efficiency may bigger than zero. The highest efficiency appears when $\eta_E=1$ and $\eta_B=0$. When $\eta_B<1$, the security of information transmission is guaranteed at the expense of encoding efficiency. This process is similar to the privacy amplification in the QKD, in which more qubit Eve obtained, less secure qubits can be left.

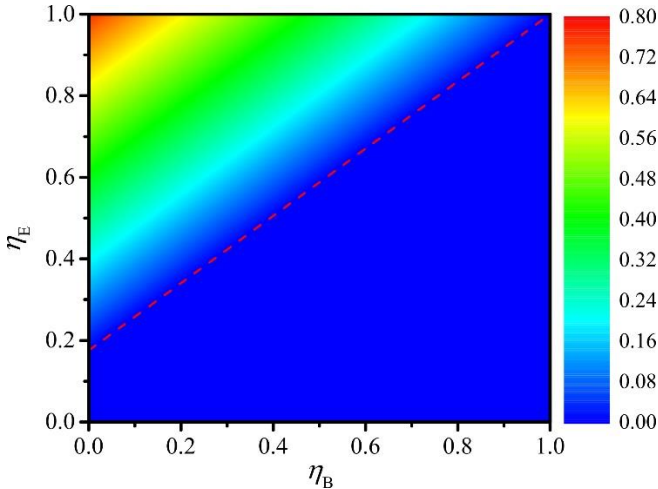

FIG. 1. The The red dash line indicates the boundary of secure and unsecure area, with the increase of $\eta_E$, the secure transmission rate increases, however, the maximum transmission rate is always smaller than one, this is because of channel noise, in this calculation we set $p_A=0.01$, $p_c=0.03$.

## III. CONCLUSION

In this paper, the security of the two-way QSDC protocol is proved against general attacks. The proof included two steps: firstly, we proved that when the system noise is lower than the preset threshold, the mutual information between Alice and Bob is larger than the mutual information shared between Alice and the eavesdropper. Secondly, once the first step is in the affirmative, one can always find a proper FEC to make the information transmission securely. Our work may be extended to other QSDC protocols, such as two-step protocol, multi-step protocol, high-dimension protocol, and so on.

## ACKNOWLEDGMENTS

We are grateful to professor Guilu Long for helpful discussions. The project is sponsored by the Natural Science Foundation of China (Nos. 61527824, 11374196, and 61675119) and PCSIRT (No. IRT 13076).